\documentclass [12pt] {article}
\usepackage{a4}

\usepackage{amsfonts}
\usepackage{amsmath}
\usepackage{graphics}
\usepackage{epsfig}

\begin{document}

\newcommand {\eps}{\varepsilon}
\newcommand {\ti}{\tilde}
\newcommand {\D}{\Delta}
\newcommand {\G}{\Gamma}
\newcommand {\de}{\delta}
\newcommand {\al}{\alpha}
\newcommand {\la}{\lambda}
\newcommand {\mO}{\mathcal{O}}
\thispagestyle{empty}

\noindent hep-th/0209184   \hfill  November 2002 \\

\noindent
\vskip3.3cm
\begin{center}
 
{\Large\bf Fractional BPS Multi-Trace Fields of
$\boldsymbol{\mathcal{N}\!=\!4}$ SYM$_{\boldsymbol{4}}$ from AdS/CFT}
\bigskip\bigskip\bigskip
 
{\large Thorsten Leonhardt, Ahmed Meziane and Werner R\"uhl}
\medskip
 
{\small\it Department of Physics\\
     Erwin Schr\"odinger Stra\ss e \\
     University of Kaiserslautern, Postfach 3049}\\
{\small\it 67653 Kaiserslautern, Germany}\\
\medskip
{\small\tt tleon, meziane, ruehl@physik.uni-kl.de}
\end{center}
 
\bigskip 
\begin{center}
{\sc Abstract}
\end{center}
\noindent
We prove inductively that every $k$-trace operator of $SO(6)_R$ irrep
with Young tableau partition $\{r_1,r_2,r_3\}$, constructed out of $k$
chiral primaries in the twenty dimensional $SO(6)_R$ irrep, leads to a
quasi primary field with protected conformal dimension. Our argument
is based on perturbative evaluations of certain four point functions
up to order $1/N^2$.

\newpage

\section{Introduction}

The Maldacena conjecture
\cite{Maldacena:1997re,Gubser:1998bc,Witten:1998qj} has opened a way
of finding nontrivial statements about certain conformal field
theories. Its actually most fruitful example is the duality between
$\mathcal{N}\!=\!4$ supersymmetric Yang Mills theory in four
dimensions and type II B string theory in the AdS$_5 \times S^5$
background. Since in this highly supersymmetric gauge theory one has a
free parameter, namely the gauge coupling constant, one can compare
results from the weak coupling regime, which can be treated via
perturbative techniques, with those from the strong coupling
regime. The latter is accessible via the AdS/CFT duality. This will also
be the philosophy of the current work, where we use the AdS/CFT
correspondence to show that certain anomalous dimensions vanish at
strong coupling.
 
In $\mathcal{N} \! = \! 4$ superconformal Yang-Mills theory SYM$_4$ an
infinite number of chiral primary fields
\begin{equation} \label{1.1}
\mathcal{O}^I_m(x)= tr_{SU(N)}
\{\Phi^{(i_1}\Phi^{i_2}\cdots\Phi^{i_m)}-\text{traces}\}
\end{equation}
exists, which is formed from the scalar fields $\Phi$ in the gauge
supermultiplet by a trace over the $SU(N)$ gauge labels. The
superscripts $\{i\}$ denote the components of an $SO(6)_R$ vector and
the rank $m$ symmetric tensor is made traceless by explicit
subtraction. In Dynkin labels\footnote{We use brackets for
$SU(4)_R$ Dynkin labels.} of $SU(4)$ the representation (\ref{1.1}) is
[0,m,0]  and its dimension is
\begin{equation} \label{1.2}
D_m = \frac{1}{12}(m+1)(m+2)^{2}(m+3).
\end{equation}
It is well known that the chiral primary fields (\ref{1.1}) belong to short
multiplets and that their conformal dimensions are protected

\begin{equation} \label{1.3}
\Delta(\mathcal{O}^I_m) = m.
\end{equation}

In this work we consider composite fields which are produced from $k$
fields $\mathcal{O}_2(x)$ by a certain stepwise fusion, see section 2. In
general these composite fields are spacetime tensors of rank $l$, they
have approximate dimension
\begin{equation} \label{1.4}
\Delta = 2k+l,
\end{equation}
and are $SO(6)_R$ traceless tensors, which we want to describe by
$SO(6)_R$ Young tableaus. Each such tableau has at most three rows of
length $ r_{1}\geq r_{2}\geq r_{3} \geq 0$ , $\{r_{1}, r_{2},
r_{3}\}$. We want to exclude contractions of $SO(6)_R$ vector labels
throughout, such that
\begin{equation} \label{1.5}
r_{1} + r_{2} + r_{3}= 2k,
\end{equation}
and $\{r_{1}, r_{2}, r_{3}\}$ is a partition of $2k$. Our fusion
procedure produces a tower of quasi-primary (or mixtures of
quasi-primary ) fields with increasing $l$. The minimal $l$ is $l_0$
\begin{equation} \label{1.6}
l_{0} =
   \begin{cases}
         0 & \text{if $ r_{1},r_{2},r_{3}$ are all even}; \\
         1 & \text {if only one $r_{i}$ is even}. \\
         \end{cases}
\end{equation}
It is important to point out that the composite fields with $l_0=0$
cannot be supersymmetric descendants, since the minimal dimension for
operators composed of $k$ operators $\mathcal{O}_2$ is reached by the $l_0=0$
fields due to (\ref{1.4}), and each supersymmetry charge raises the
dimension by $1/2$. On the other hand, with the help of
\cite{Dolan:2002zh} one can show that the composite fields with
$l_0=1$ can be obtained by the application of two (different)
supersymmetry charges from a composite field with the same number of
$SO(6)_R$ blocks. Thanks to supersymmetry the composite fields with
$l_0=1$ inherit their properties from those with $l_0=0$.

The representation of $SO(6)_R$ characterized by the partition
$\{r_{1}, r_{2}, r_{3}\}$ is reducible if $r_{3}> 0$. It decomposes
into a conjugate pair of a self-dual and an antiself-dual
representation with Dynkin labels 
\begin{equation} \label{1.7}
[r_{2}- r_{3}, r_{1} - r_{2}, r_{2} + r_{3}]\oplus[r_{2}+ r_{3},
r_{1}- r_{2}, r_{2} - r_{3}].
\end{equation}
We will show that in the case $l = l_0$ these fields have protected
dimensions $2n+l_0$. The scalars ($l_0=0$) probably belong to
$\frac{1}{8}$-BPS multiplets, whereas the vectors as supersymmetric
descendants might be $\frac{1}{4}$-BPS or $\frac{1}{2}$-BPS fields as
well. If $r_{3}= 0$ then the Dynkin labels reduce to
\begin{equation} \label{1.8}
[r_{2},r_{1}- r_{2},r_{2}].
\end{equation}
We find that the corresponding field with $l = l_0$ has a protected
dimension and the scalar supposedly belongs to a $\frac{1}{4}$-BPS
multiplet. Again, the BPS-type of the vector can only be determined in
a case by case consideration. Finally for $ r_{2} = r_{3} = 0$ we are
led to
\begin{equation} \label{1.9}
[0,r_{1},0] \;\;\;\;\; (\text{$r_{1}$ even} );
\end{equation}
which is the case dealt with by Skiba \cite{Skiba} if $l = 0$ and
belongs to $\frac{1}{2}$-BPS multiplets presumably. Our proof of
protectedness is based on AdS/CFT correspondence and a perturbative
expansion in $1/N^2$ and takes account only of the first
order $O(\frac{1}{N^2}) $.

In our construction the number of $SU(N)$-traces is maximal. Other
authors \cite{Ryzhov:2001bp,D'Hoker:2001bq} have investigated similar
problems with fields containing two and three traces only, and then
resolved the mixtures of the quasi-primary fields algebraically or
numerically. Instead we assume that for $l = l_0$ we obtain pure
quasiprimary fields. This assumption is based on the fact that in free
field theory there is no freedom to place one ($l_0 = 1$) or no ($l_{0}=
0$) derivatives into a normal product. However, for $l > l_0$ there is
always such freedom and mixtures must appear.

Further results on vanishing anomalous dimensions have been found in
\cite{Eden:2001ec,Ferrara:2001uj}, see also \cite{Heslop:2001gp}, 
by an analysis of three-point functions of two $\frac{1}{2}$-BPS
operators with any other possible operator in a purely group theoretic
fashion.\footnote{We thank E.~Sokatchev for bringing our attention to
this point.} This kinematical approach nicely complements our results
in two ways: On the one hand our approach by considering four point
functions is based on a $1/N$-expansion of certain four point
functions and is thus dynamical. On the other hand, in the references
given above, protected dimensions for operators with $SU(4)_R$ irreps
contained in the reduction of $[0,p,0]\otimes[0,q,0]$ for any $p,q$
are obtained, whereas we only have results for even
$p,q$. Nevertheless, we find vanishing anomalous dimensions for fields
with $SU(4)_R$ irreps which are not contained in the reduction of any
$[0,p,0]\otimes[0,q,0]$, but which are contained in a chain of
reductions. E.g. we get a protected dimension for a field with
$SU(4)_R$ irrep $[0,2,4]$, which cannot be obtained by a single
reduction from the above factors, but which is contained in the
``triple'' reduction
\begin{multline}
\Bigl(\Bigl([0,2,0]\otimes[0,2,0]\Bigr)\otimes[0,2,0] \Bigr) \oplus
[0,2,0] \\ 
= \Bigl(\Bigl([2,0,2]\oplus\cdots \Bigr)\otimes[0,2,0]\Bigr) \oplus
[0,2,0]  \\  
= \Bigl([2,2,2]\oplus\cdots \Bigr) \otimes [0,2,0] = [0,2,4] \oplus
\cdots. 
\end{multline}
  

\section{The fusion procedure}

\begin{figure}[htb]
 \begin{centering}
\resizebox{10cm}{!}{
\includegraphics{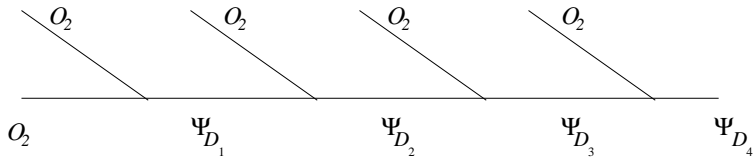}}
\caption{The fusion procedure}\label{fusion}
 \end{centering}
\end{figure}

The general idea of our method to produce multitrace operators with
protected dimensions is to consider a stepwise fusion process, which
may be sketched by fig. \ref{fusion}.

In this picture each vertex denotes a fusion of the two incoming
fields with protected dimensions to the outgoing one. By fusion we mean the
formation of the normal product according to some rules given
below and, since the normal product of two fields is in general
reducible under $SO(6)_R$, we have to project onto an $SO(6)_R$ irrep.

The operators $\mO_2$ are the single trace operators
\begin{align}
\mO_2^I (x) = tr_{SU(N)} \{ \Phi^{(i} \Phi^{j)} - \frac{1}{6} \Phi^2
\de^{ij}\} , 
\end{align}
where $i,j$ are $SO(6)_R$ vector labels and $I$ denotes a basis of the
twenty dimensional representation space of the symmetric traceless
irrep of $SO(6)_R$ under which $\mO_2^I$ transforms. It is well known
that $\mO_2^I$ as chiral primary operator has protected dimension $2$
and we normalize its two-point function as
\begin{align}\label{twoptfct}
\langle \mO_2^{I_1}(x_1) \mO_2^{I_2}(x_2) \rangle = \frac{\de^{I_1
I_2}} {(x_{12}^2)^2}.
\end{align}

The operators $\Psi_{D_i}$ are multitrace operators, which appear in
the normal product of the two incoming operators $\mO_2$ and
$\Psi_{D_{i-1}}$. These fields transform in the irrep $D_i$ of
$SO(6)_R$.  

The rules for the fusion procedure are:
\begin{enumerate}
\item Discard $SO(6)_R$ representations with contractions and thereby
all singular terms in operator product expansions (``block number
conservation'').
\item Admit only scalar intermediate quasi-primary fields.
\end{enumerate}

We will show that any irrep appearing in this sequence of fusions
contains at least one field with protected dimension and that thus
the dimension of this field equals the sum of the dimensions of its
factors. 

We do not obtain every $SO(6)_R$ irrep if we start only with scalar
factors. The set of fields,which are out of range of our fusion
procedure with only scalar fields in the input, consists of fields
with the following $SO(6)_R$ irreps:
\begin{align}\label{except}
\{2r+1,2r+1,2s\} \; \textrm{with}\; r \ge s, \nonumber \\
\{2s,2r+1,2r+1\} \; \textrm{with}\; r < s.
\end{align}
But since these fields are spacetime vectors according to (\ref{1.6}),
they are supersymmetric descendants of fields with $l_0=0$. Since our
proof holds for the $l_0=0$ fields, we can conclude that the fields
with $SO(6)_R$ irrep (\ref{except}) also have protected conformal
dimension.


\subsection{The first and simplest example}

Besides being illustrative, the first step of the fusion procedure is
the start of our inductive proof, so we repeat the arguments contained
in \cite{Arutyunov:2000ku,Eden:2001ec,Dolan:2001tt,Arutyunov:2001mh},
which are useful for us. The $SO(6)_R$ representation of the operator
product $\mO_2^{I_1}(x_1) \mO_2^{I_2}(x_2)$ decomposes into irreps
\begin{align}\label{zerl_22}
\{2,0,0\} \otimes \{2,0,0\} = & \{4,0,0\} \oplus \{3,1,0\} \oplus
\{2,2,0\} \nonumber \\
& \oplus \{2,0,0\} \oplus \{1,1,0\} \oplus \{0,0,0\}.
\end{align} 
The irreps in the second line contain at least one contraction and are
discarded according to rule 1. Then the four-point function $\langle
\mO_2^{I_1}(x_1) \mO_2^{I_2}(x_2) \mO_2^{I_3}(x_3)
\mO_2^{I_4}(x_4)\rangle$ can be analyzed in terms of conformal partial
waves, where each partial wave transforms under $SO(6)_R$ in one of
the irreps of the right hand side of (\ref{zerl_22}). This four-point
function is calculated in \cite{Arutyunov:2000ku} up to order
$O(1/N^2)$ at strong coupling by an AdS/CFT calculation. The result
has the following form 
\begin{multline}\label{Aru4pt}
\langle \mO_2^{I_1}(x_1) \mO_2^{I_2}(x_2) \mO_2^{J_1}(x_3)
\mO_2^{J_2}(x_4)\rangle  \\
= (x_{13}^2 x_{24}^2)^{-2} u^{-2} \Biggl\{\sum_{r=0}^2 \Bigl[1 +(-1)^r
\bigl(1-Y \bigr)^2 \Bigr] P_{\{4-r,r,0\}}^{I_1 I_2,J_1 J_2}  \\
+ \frac{1}{N^2} \sum_{r=0}^2 \Bigl[ \la_{\{4-r,r,0\}}(u,Y) \phi (u,v)
+ \mu_{\{4-r,r,0\}}(u,Y) \Bigr] P_{\{4-r,r,0\}}^{I_1 I_2,J_1 J_2} \\ +
O(\frac{1}{N^4}) + \textrm{discarded irreps}\Biggr\},
\end{multline}
where we used the biharmonic conformal invariants
\begin{align}
u=\frac{x_{12}^2 x_{34}^2}{x_{13}^2 x_{24}^2},\; v=\frac{x_{14}^2
x_{23}^2}{x_{13}^2 x_{24}^2}, \;Y=1-\frac{1}{v}, 
\end{align}
which are suited for an operator product expansion of (\ref{Aru4pt}),
since in this case we have 
\begin{align}
x_{12} \to 0,\; x_{34} \to 0
\quad \Leftrightarrow \quad u \to 0, \; v \to 1, \; Y \to 0.
\end{align} 
The representation dependent functions $ \la_{\{4-r,r,0\}}(u,Y)$ are
given by  
\begin{align}\label{laofr}
\la_{\{4,0,0\}}(u,Y) & = u^2(1-Y), \nonumber \\
\la_{\{3,1,0\}}(u,Y) & = u Y(1-Y), \nonumber \\ 
\la_{\{2,2,0\}}(u,Y) & = \frac{3}{2}u(1-Y)(2-Y)-\frac{1}{2}
u^2(1-Y)^2,
\end{align}
as well as
\begin{align}\label{muofr}
\mu_{\{4,0,0\}}(u,Y) & = 4 u^2 (1-Y) \nonumber \\
\mu_{\{3,1,0\}}(u,Y) & = 0 \nonumber \\ 
\mu_{\{2,2,0\}}(u,Y) & = -2 u^2 (1-Y).
\end{align}
Moreover we need a generalized hypergeometric function
\cite{Hoffmann:2001gk,Eden:2000bk} 
\begin{multline} \label{phiofu}
\phi(u,v)= -4 \, \frac{1}{1-Y} \frac{\partial}{\partial \eps}\biggr
\rvert_{\eps=0} \sum_{n\ge0} \frac{u^{n+2+\eps}}{\G(n+1+\eps)}
\frac{\G(n+3+\eps)^2 \,\G(n+4+\eps)}{\G(2n+6+2\eps)} \\
  F \Bigl[{n+3+\eps, n+2+\eps \atop 2n+6+2\eps }; 1-v \Bigl] 
\end{multline}

To obtain the contribution of an exchanged field, which transforms in
an irrep $\{4-r,r,0\}, \,r=0,1,2$, of the first line of the right hand
side of (\ref{zerl_22}), we have to project onto this irrep by the
respective projection operator $P_{\{4-r,r,0\}}$. Since these
projectors are mutually orthogonal, application of $P_{\{4-r,r,0\}}$
to (\ref{Aru4pt}) gives only terms proportional to $P_{\{4-r,r,0\}}$
 on the right hand side.

Note that the logarithmic terms, which appear at $1/N^2$, have the
form (supressing indices)
\begin{align}\label{mainpoint}
P_{\{4-r,r,0\}}\langle
\mathcal{O}\mathcal{O}\mathcal{O}\mathcal{O}\rangle \Bigr \rvert_{\sim
\log} \sim \log u \bigl(\, u^p + \textrm{higher powers in}\; u \bigr),
\end{align}
with $p=2$ if $r=0,2$ and $p=1$ if $r=1$.

The general form of the exchange of a partial wave with dimension
$\de$ and spacetime tensor rank $l$ produced from two
scalar fields of dimensions $\D _1$ and $\D _2$ can be found with the
help of the ``master formula'' of \cite{Lang:1992zw}
\begin{align} \label{master} 
 (x_{13}^2)^{-\D_1} (x_{24}^2)^{-\D_2} C_{l,\D_1,\D_2} \sum_{M=0}^l
u^{\frac{1}{2}(\de -\D_1 -\D_2 -M)} f_M(u,1-v), 
\end{align}
where $f_M$ is an analytic function in $(u,1-v)$ with $f_M(0,1-v)\neq 0$,
and $C_{l,\D_1,\D_2}$ is a constant containing the
normalization of the exchanged field and the coupling of the
external fields to the exchanged one. The dimension $\de$ of the
exchanged composite operator built out of the components with
dimensions $\D_1$ and $\D_2$ is given by
\begin{align}
\de = \D_1+\D_2+l+2t+\eta_{l,t}.
\end{align}
Plugging this formula into (\ref{master}) and expanding
$u^{\eta_{l,t}}= 1+\eta_{l,t}\log u +\cdots$ we see that the leading
term proportional to $\log u$ begins with the constant
$\eta_{l_0,0}$. To achieve agreement with the strong coupling result
(\ref{mainpoint}), $\eta_{l_0,0}$ must be zero at strong coupling and
we conclude that $\eta_{l_0,0}$ vanishes at all.

Now we take a closer look at the four point function leading to the
exchange of the field with $SO(6)_R$ irrep $\{3,1,0\}$ and observe
that it has a factor $Y$. The Taylor expansion of $Y$ in terms of
$x_{12}^\mu, x_{34}^\nu$ starts with
\begin{align}
\frac{2}{x_{24}^2} t_{\mu \nu}(x_{24}) x_{12}^\mu x_{34}^\nu.
\end{align}
On the other hand, the numerator in the two-point function of a
conformal vector field contains the tensor
\begin{align}
t_{\mu \nu}(x) = \de_{\mu \nu} - 2 \frac{x_\mu x_\nu}{x^2}.
\end{align}
This indicates that the exchanged field with $SO(6)_R$ irrep
$\{3,1,0\}$ is indeed a vector field, in agreement with (\ref{1.6}).


\subsection{The second example}

We now consider the fusion $\Psi_{\{2k-2,0,0\}}^{I_1 I_2 \ldots I_{k-1}}
(x_1) \mO_2^{I_{k}} (x_2)$, which was also done in
\cite{Hoffmann:2001gk}. According to Skiba's theorem \cite{Skiba}, the
scalar field $\Psi_{\{2k-2,0,0\}}$ has protected conformal dimension
and this dimension therefore equals $2k-2$. We want to show that the
fusion products also have protected dimension. To this end we again
consider the four-point function
\begin{align}\label{fourpt2k}
\biggl \langle \Psi_{\{2k-2,0,0\}}^{I_1 I_2 \ldots I_{k-1}} (x_1)
\mO_2^{I_{k}} (x_2) \Psi_{\{2k-2,0,0\}}^{J_1 J_2 \ldots J_{k-1}} (x_3)
\mO_2^{J_{k}} (x_4) \biggr\rangle.
\end{align}
This four-point function is calculated up to order $O(1/N^2)$ by
inserting the four-point function (\ref{Aru4pt}) into the $2k$ point
function
\begin{align}\label{twok2pt}
\biggl\langle \prod_{r=1}^{k-1} \mO_2^{I_r} (x_1^{(r)}) \mO_2^{I_{k}}
(x_2) \prod_{r=1}^{k-1} \mO_2^{J_r} (x_3^{(r)}) \mO_2^{J_{k}} (x_4)
\biggr \rangle 
\end{align}
in all possible ways.
\begin{figure}[htb]
 \begin{centering}
\resizebox{5cm}{4cm}{
\includegraphics{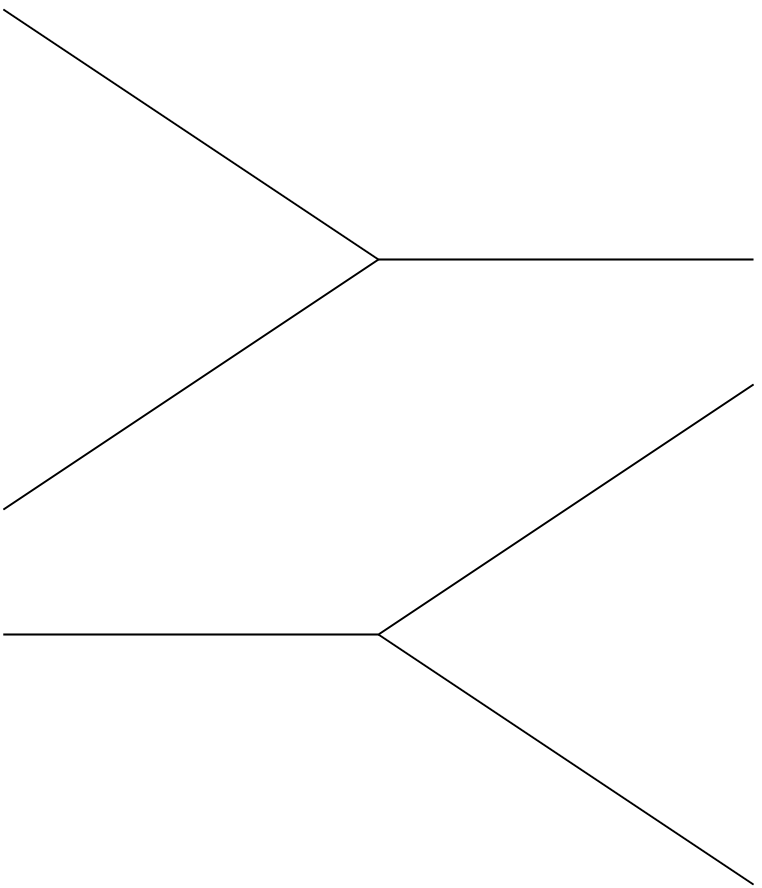}}
\resizebox{5cm}{4cm}{
\includegraphics{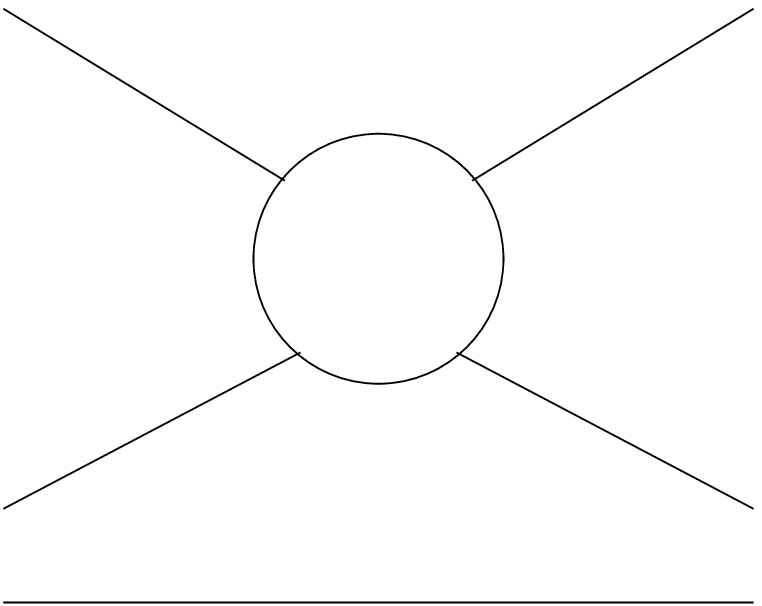}}
\caption{Contributions to the six-point functions of order
$O(1/N^2)$}\label{orderNminus2}
 \end{centering}
\end{figure}
In principle it is possible that there are also contributions of two
three point correlators, which are of order $O(1/N)$ each, to the 2k
point function for $k\ge3$, see fig. \ref{orderNminus2}. However, it is
pointed out in \cite{Hoffmann:2001gk} that discarding of $SO(6)_R$
contractions eliminates these contributions in calculations up to
order $O(1/N^2)$.

The next step in obtaining the four-point function (\ref{fourpt2k}) is
the projection of $2k-2$ $SO(6)_R$ indices of the $2k$ point function
(\ref{twok2pt}) onto the $\{2k-2,0,0\}$ irrep. Then we perform the
``local limit'' $x_i^{(r)} \to x_i$ to obtain the four-point function
(\ref{fourpt2k}).  Finally we project onto the irreps which occur in
the normal product $\Psi_{\{2k-2,0,0\}}^{I_1 I_2 \ldots I_{k-1}}
(x_1)$ $\mO_2^{I_{k}} (x_2)$, i.e.
\begin{align}
\{2k-2,0,0\} \otimes \{2,0,0\} & =  \{2k,0,0\} \oplus \{2k-1,1,0\} \oplus
\{2k-2,2,0\} \nonumber \\
& \qquad +\textrm{irreps with contractions}.
\end{align}

The result of this computation found in \cite{Hoffmann:2001gk} yields
for any irrep $\{2k-r,r,0\}$, $r\in\{0,1,2\}$,
\begin{align}\label{reshogk}
\biggl \langle &\Psi_{\{2k-2,0,0\}}^{I_1 I_2 \ldots I_{k-1}} (x_1)
\mO_2^{I_{k}} (x_2) \Psi_{\{2k-2,0,0\}}^{J_1 J_2 \ldots J_{k-1}} (x_3)
\mO_2^{J_{k}} (x_4) \biggr\rangle = (x_{13}^2)^{-2k+2}
(x_{24}^2)^{-2} (k-1)! \nonumber \\
& \Biggl\{ \biggl[ a(k,r) + b(k,r) \bigl(1-Y\bigr)^2 \biggr] \nonumber
\\
& \quad + \frac{1}{N^2} \biggl[ c(k,r) + d(k,r)\bigl(1-Y\bigr)^2
\nonumber \\
& \qquad + e(k,r)\bigl(1-Y\bigr)^2 \Bigl[ \la_{\{4-r,r,0\}}(u,Y) \phi
(u,v) + \mu_{\{4-r,r,0\}}(u,Y) \Bigr] \biggr] \nonumber \\ & \qquad
\qquad \qquad \qquad \qquad
+ O(1/N^4) \Biggr\} P_{\{2k-r,r,0\}}^{I_1 \cdots I_k, J_1 \cdots
J_k}+\textrm{irreps with contractions}. 
\end{align}
 The functions $\mu, \la, \phi$ are the same as in
(\ref{laofr}),(\ref{muofr}),(\ref{phiofu})
and the coefficient functions $a,b,c,d,e$ are given in table \ref{Tabell}.

\begin{table}[htb]
\caption[Tabelle]{Table of coefficients $a$-$e$}\label{Tabell}
\begin{center}
\begin{tabular}{c|c|c|c|c|c}
 & $a$ & $b$ & $c$ & $d$ & $e$ \\ \hline
$\{2k,0,0\}$ & $ 1 $ & $k-1$ & $(k-1)(k-2)$ &
$(k-1)(k-2)(k+1)$ & $k-1 $\\
$\{2k-1,1,0\}$ & $ 1 $ & $-1$ & $(k-1)(k-2)$ &
$-(k-1)(k-2)$ & $k-1 $\\
$\{2k-2,2,0\}$ & $ 1 $ & $k-1$ & $(k-1)(k-2)$ &
$\frac{2}{3}(k-1)(k-2)$ & $k-1 $ \\ \hline
\end{tabular}
\end{center}
\end{table}
By the general rule (\ref{1.6}), in the case $r=1$ we expect that the
exchanged partial wave of lowest spacetime tensor rank is a
vector. Therefore we must have proportionality with $Y$. This is in
fact true since from the second line of table \ref{Tabell} we read off
that
\begin{align}
a(k,1) + b(k,1) & = 0 \nonumber \\
c(k,1) + d(k,1) & = 0,
\end{align}
and $\la_{\{3,1,0\}}(u,Y)$ and $\mu_{\{3,1,0\}}(u,Y)$ allow to factor
$Y$, see (\ref{laofr}),(\ref{muofr}).

Since the terms proportional to $\log u$ all stem from the $O(1/N^2)$
contributions of the four point function, they are at least of order
$u$, and comparison with (\ref{master}) reveals again that the
anomalous dimension of the composite field with minimal dimension
vanishes. More precisely, in the case $\{2k,0,0\}$ the dimensions
$\de=2k, 2k+2$ do not acquire an anomalous term, while for the other
two cases only the fields with minimal conformal dimension $\de=2k,
\de=2k+1$ for the scalar and the vector, respectively, have vanishing
anomalous dimensions.

We mention finally that the effective perturbation expansion parameter
in (\ref{reshogk}) is obviously $\frac{k}{N}$ and not $\frac{1}{N}$,
which must be much smaller than one to enable such an expansion. From
the critical nonlinear $O(N)$ sigma model a similar behavior is known,
where the effective expansion parameter is $\frac{k}{N^{1/2}}$
\cite{Lang:1993ct}, where $k$ denotes the number of fusions of
fundamental fields.


\section{The general case}

For the general case we consider the fusion of a scalar normal product
$\Psi_{\{r_1,r_2,r_3\}}$, where $r_i$ are all even, of protected conformal
dimension $\D = 2k-2 = r_1+r_2+r_3$ with $\mO_2$. We claim that to
any irrep of $SO(6)_R$ appearing in the reduction of $\{r_1,r_2,r_3\}
\otimes \{2,0,0\}$ respecting the fusion rules belongs at least one
field with vanishing anomalous dimension. The proof goes along similar
lines as before.

\begin{figure}[htb]
\begin{centering}
\resizebox{6cm}{!}{
\includegraphics{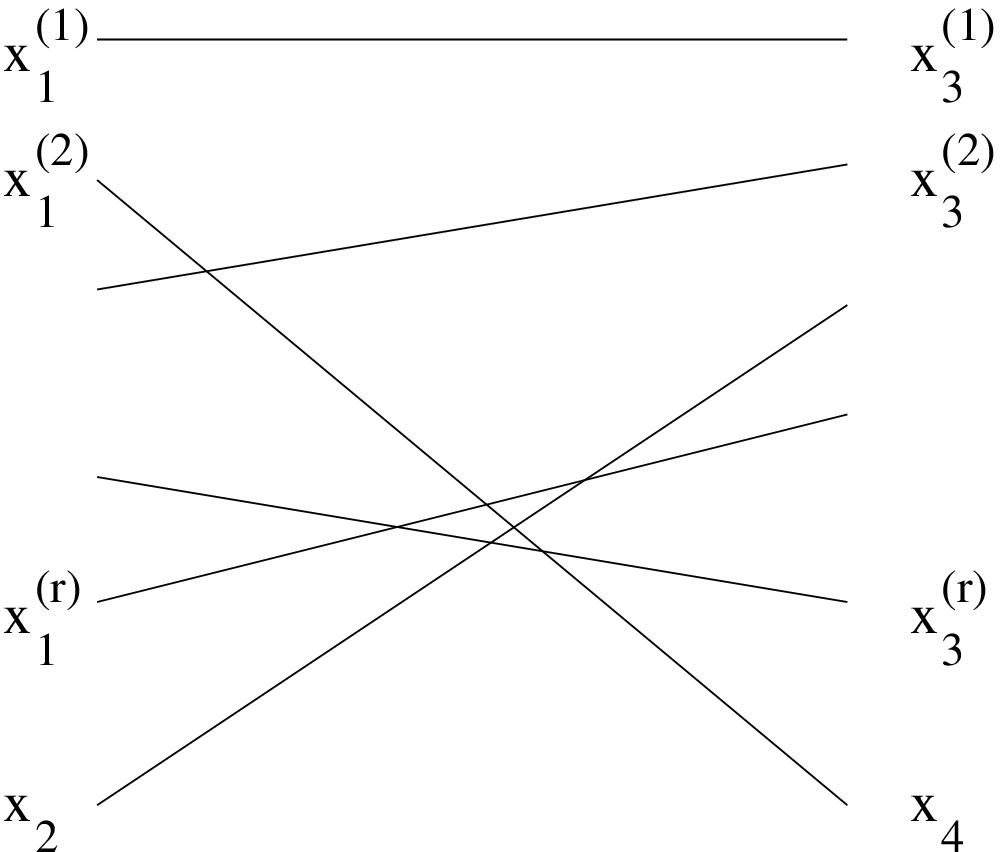}}
\caption{$O(1)$ contribution: The $2k$ point function decomposes into
a product of two-point functions }\label{O1con}
\end{centering}
\end{figure}

\begin{figure}[h]
\begin{centering}
\resizebox{6cm}{!}{
\includegraphics{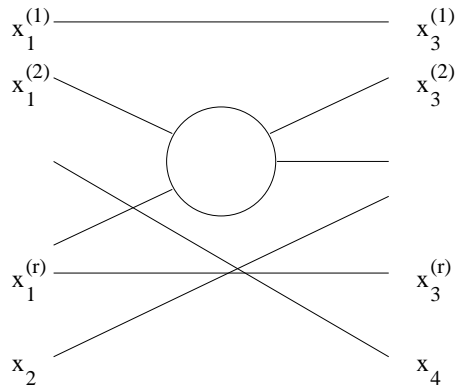}}
\caption{Class (1) graphs where both pairs of legs coincide.}\label{class1}
\end{centering}
\end{figure}

\begin{figure}[h]
\begin{centering}
\resizebox{6cm}{!}{
\includegraphics{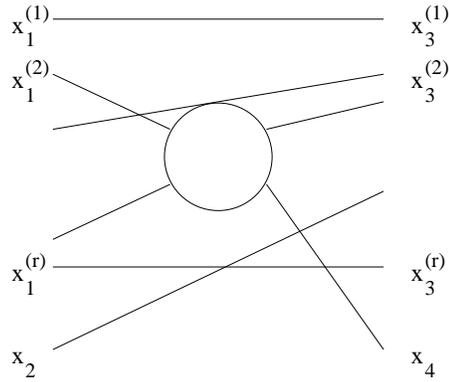} }
\caption{Class (2) graphs where the left pair of legs of the
four-point function insertion coincide. The class (3) graphs are
analogous to the class (2) graphs, with the right pair of legs of the
inserted four point function coinciding.}\label{class2}
\end{centering}
\end{figure}

\begin{figure}[h]
\begin{centering}
\resizebox{6cm}{!}{
\includegraphics{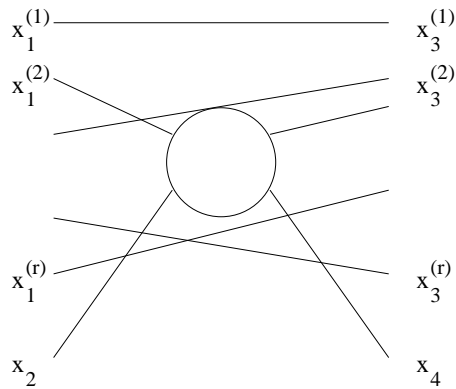}}
\caption{Class (4) graphs: no legs of the inserted four point function
coincide }\label{class4}
\end{centering}
\end{figure}

First we calculate the four-point function
\begin{align}\label{gen4pt}
\biggl\langle \Psi_{\{r_1,r_2,r_3\}}^{I_1 \ldots I_{k-1}}(x_1)
\mO_2^{I_{k}}(x_2) \Psi_{\{r_1,r_2,r_3\}}^{J_1 \ldots J_{k-1}}(x_3)
\mO_2^{J_{k}} (x_4) \biggr \rangle
\end{align}
by evaluating the $2k$ point function
\begin{align}
\biggl\langle \prod_{r=1}^{k-1} \mO_2^{I_r} (x_1^{(r)}) \mO_2^{I_{k}}
(x_2) \prod_{r=1}^{k-1} \mO_2^{J_r} (x_3^{(r)}) \mO_2^{J_{k}} (x_4)
\biggr \rangle 
\end{align}
up to order $O(1/N^2)$. In this calculation we are confronted with the
following graphs:
\begin{itemize}
\item At order $O(1)$ the $2k$ point function decomposes into
a product of two-point functions (see fig. \ref{O1con}).

\item At order $O(1/N^2)$ we have to distribute respectively
two legs of the $O(1/N^2)$ part of the four point function
(\ref{Aru4pt}) to the external points $x_1^{(r)},x_2$ and
$x_3^{(r)},x_4$ in all possible ways. Then we must connect the
residing external points via two point functions. This leads to four
classes of graphs shown in figures \ref{class1}, \ref{class2},
\ref{class4}.

\end{itemize}


Then we sum up the $k!$ graphs from $O(1)$ and the
$\binom{k}{2}^2(k-2)!$ graphs from $O(1/N^2)$. To be able to perform
the local limit $x_i^{(r)} \to x_i$ we must first project onto the incoming
irreps. In this limit we then obtain the four-point function
(\ref{gen4pt}).

In the following step we project onto the block-conserving irreps
$\{r'_1,r'_2,r'_3\}$, which occur in the reduction of $\{r_1,r_2,r_3\}
\otimes \{2,0,0\}$.  Thus we need the following projections:
\begin{itemize}
\item At order $O(1)$
\begin{align}
P_{\{r'_1,r'_2,r'_3\}\; K_1 \ldots K_{k-1} I_k}^{J_1 \ldots J_k}
P_{\{r_1,r_2,r_3\}\;\, I_1 \ldots I_{k-1}}^{K_1 \ldots K_{k-1}}
\prod_{m=1}^{k} P_{\{2,0,0\}\; J_m}^{I_m},
\end{align}
\item and at order $O(1/N^2)$, say for class (4), see fig. \ref{class4}
\begin{multline}
P_{\{r'_1,r'_2,r'_3\}\;K_1 \ldots K_{k-1} I_k}^{L_1 \ldots L_{k-1}
J_k} P_{\{r_1,r_2,r_3\}\; I_1 \ldots I_{k-1}}^{K_1 \ldots K_{k-1}}
\prod_{m=1}^{k-2} P_{\{2,0,0\}\; J_m}^{I_m} P_{\{4-r,r,0\} \; J_{k-1}
J_k}^{I_{k-1} I_k} \\ P_{\{r_1,r_2,r_3\}\; L_1 \ldots L_{k-1}}^{J_1
\ldots J_{k-1}}.
\end{multline}
\end{itemize}
To avoid the evaluation of these projectors we make an ansatz for the
four-point function
\begin{align}\label{ansatz}
& P_{\{r'_1,r'_2,r'_3\}\; I_1 \ldots I_k}^{L_1 \ldots L_{k}}
P_{\{r'_1,r'_2,r'_3\}\; J_1 \ldots J_k}^{K_1 \ldots K_{k}}
\biggl\langle \Psi_{\{r_1,r_2,r_3\}}^{I_1 \ldots I_{k-1}}(x_1)
\mO_2^{I_{k}}(x_2) \Psi_{\{r_1,r_2,r_3\}}^{J_1 \ldots J_{k-1}}(x_3)
\mO_2^{J_{k}} (x_4) \biggr \rangle \nonumber \\
& = (x_{13}^2)^{-2k+2} (x_{24}^2)^{-2} \Biggl\{ \biggl[ A + B
\bigl(1-Y\bigr)^2 \biggr] \nonumber \\
& \quad + \frac{1}{N^2} \biggl[ C +
D\bigl(1-Y\bigr)^2
\nonumber \\
& \qquad \qquad + E \bigl(1-Y\bigr)^2 \Bigl[
\la_{\{4-r,r,0\}}(u,Y) \phi (u,v) + \mu_{\{4-r,r,0\}}(u,Y) \Bigr]
\biggr] \nonumber \\ & \qquad \qquad \qquad \qquad \qquad \qquad
\qquad \qquad \qquad \qquad + O(1/N^4) \Biggr\}
P_{\{r'_1,r'_2,r'_3\}}^{L_1 \cdots L_k,\, K_1 \cdots K_k}.
\end{align}
The functions $A,B,C,D,E$ are polynomials resulting from the
combinatorics of the graphs contributing to this four point
function. They depend on the representations $\{r'_1,r'_2,r'_3\}$ and
on $\{r_1,r_2,r_3\}$. 
In the local limit two (four) external legs of the four point function
eventually coincide, as one can see from the graphs of class
(1), (2) and (3). This means that we must evaluate the bracket
containing $\la \phi +\mu$ at $u=Y=0$, which gives rise to the $C$ and
$D$ function. The factor $(1-Y)^2$ arises from crossing, namely a
propagator $(x_{23}^2)^{-2}$ or $(x_{14}^2)^{-2}$. In the case of
$l_0=1$ (two $r'_i$ are odd and the third is even) we have the
constraints
\begin{align}
A + B & = 0, \nonumber \\
C + D & = 0.
\end{align}

Finally the same argument for the protectedness of the conformal
dimension of quasiprimary fields with $SO(6)_R$ representation and
spacetime tensor rank $l=l_0\in\{0,1\}$ applies. We compare the terms
proportional to $\log u$ in the operator product expansion from
(\ref{ansatz}) with those from the generic exchange (\ref{master})
and observe that the former starts with (at least) one power of $u$,
while the latter begins with a constant. Therefore we conclude that
the fields with minimal canonical conformal dimension have vanishing
anomalous dimension, i.e. the canonical conformal dimension is exact.

This completes the proof.


\section{Summary}
In this note we described a simple method of finding quasiprimary
fields with protected conformal dimensions in $\mathcal{N}\!=\!4$
supersymmetric Yang-Mills theory in four dimensions in a recursive
way. We find vanishing anomalous dimensions for spacetime scalars as
well as spacetime vectors. We noted that the vectors are supersymmetry
descendants of spacetime scalars with the same number of
$\mathcal{O}_2$ constituents, thus inheriting the protectedness of the
conformal dimension from the scalars.


\section*{Acknowledgement}

A.~M. would like to thank the German Academic Exchange Service (DAAD)
for financial support.


\clearpage

\section*{Appendix: Graphs contributing to $2k$ point functions}

\clearpage


\end{document}